\newcommand{\tit}[1]{``#1,''}
\newcommand{\ket}[1]{| #1 \rangle}

\def\ox{\otimes}
\newcommand{\Hil}{\mathcal H}
\newcommand{\Eq}[1]{Eq.~(\ref{#1})}

\documentclass[twocolumn]{article}

\begin{document}
\sloppy

\title{Comment on ``How the No--Cloning Theorem Got its Name''}

\author{\em Alexander Yu.\ Vlasov}
\date{}

\maketitle

\begin{abstract}
A review by A. Peres \cite{cloname} appears recently. It is difficult to
add something to such kind of fundamental themes, but here is briefly
presented some ideas about challenges of the no-cloning theorem and imaginary
modifications of quantum mechanics, that could make precise cloning possible.
\end{abstract}

\section*{Introduction}

The no-cloning theorem \cite{noclon} based on idea of nonlinearity is
very straightforward and already
hint like ``quantum cloning is impossible, because it is not linear'' is
usually enough for a quantum physicist to recover at least outline of the
proof.\footnote{One negative result of such situation is following: I knew
some people ``spoiled by modern electronic technology,'' who simply did not
read yet original paper. I would recommend them after all to find a time to
see \cite{noclon}.} Why it was really not issued already 75 years ago?

I think, together with some new details uncovered in \cite{cloname}
there is omnipresent ``meta-physical'' problem --- correspondence between
a pure mathematical model and the real physical system, and here
no-cloning theorem provides some challenges, possibly reflected already in
first replies on the article \cite{noclonrep}. I would try to draw this
trouble in most general terms --- simple and clear mathematical model
used in \cite{noclon} stimulates some questions about physical principle
of {\em covariance}, even more general, than linearity of quantum mechanics.

Sure, it is possible to produce modification of no-cloning proof to take into
account this more general principle for particular model with photons
and lasers, but in such a case it loses some charm of universality
and simplicity of mathematical arguments. So, here I am briefly discussing
yet another idea --- considering imaginary modifications of quantum
evolution, like nonlinear model briefly discussed in \cite{cloname}, there
perfect cloning could be acceptable.

\section{Covariance vs. linearity?}

Let us use simplified formulations of no-cloning theorem without state of
measurement device often used nowadays. There are two basic schemes, those 
could be found in literature:
\begin{equation}
 \ket{\psi} \to \ket{\psi}\ket{\psi},
\label{clon12}
\end{equation}
and, more accurate:
\begin{equation}
 \ket{a}\ket{\psi} \to \ket{\psi}\ket{\psi},
\label{clon22}
\end{equation}
where $\ket{a}$ is some fixed, known state of ancillary system, sometime
denoted simply as $\ket{0}$ due to ``fashionable'' applications in quantum
information science.

Maybe second expression \Eq{clon22} is not ``more accurate'' and
in \cite{noclon} was used rather first setup\footnote{More precisely it was
considered also state of measurement device, but because it was noted also,
that the state is unchanged in second version of theorem, it can be suggested,
that auxiliary system like $\ket{a}$ in \Eq{clon22} was not treated as
part of device.}, but more universal and understanding arguments are
related rather with second one.

\medskip

Anyway, let us instead of proper and difficult model of first setup
\Eq{clon12} with Fock spaces, continuous variables, {\em etc.}, consider
na{\"\i}ve {\em nonstandard, ``modified''} quantum evolution between two
finite-dimensional Hilbert spaces with {\em different dimensions}:
\begin{equation}
 \Hil \to \Hil\ox\Hil.
\label{evol12}
\end{equation}

In general, relations between such spaces with different dimensions are not
always functions at all ({\em i.e.} multi-functions). It is possible to ask,
that kind of usual functions, maybe even nonlinear, for such non-standard
evolution are appropriate as
generalization of unitary linear maps between spaces with equal dimension.

Maybe natural example is unitary linear map
between first space and subspace of second space with same dimension (image
of linear map). Really it is not very interesting example, because it does
not differ much from standard case with two equal spaces --- if second
one contains subspace that never could be reached, why simply does not
cut this ``nonphysical junk''? But it is not the only problem.

Let us consider usual linear cloning of two orthogonal states allowed
by unitary quantum evolution:
\begin{equation}
\ket{0} \mapsto \ket{0}\ket{0}; \quad
\ket{1} \mapsto \ket{1}\ket{1},
\label{linclon}
\end{equation}
with arguments used in \cite{noclon} we have result,
that linear, unitary evolution may clone only this two states.

Let us consider now situation, when modified (imaginary, nonstandard)
evolution \Eq{evol12} describes decay of some hypothetical
free spin-1/2 particle into two spin-1/2 particles. For this
rather unrealistic\footnote{Main problem here is conservation of angular
momentum (if do not ask about less fundamental quantum numbers). It presents
also in initial paper \cite{noclon} and was discussed in replies
\cite{noclonrep}. But for our hypothetical process \Eq{evol12} the difficulty
with momentum is rather inherited, because already in classics conservation
laws due to Noether's theorem can be considered as consequence of space-time
symmetries and {\em homogeneity}, it has important influence also in
quantum theory \cite{QFT}, but process \Eq{evol12} may not be described
with {\em continuous} time, because dimension of phase space spasmodically
increases. So consideration below could be considered as tries to introduce
reasonable laws for this discontinuous process and seems the violation is
really minimal --- using classical analogue in discussed hypothetical model
is changed only absolute value of angular momentum, not direction, and such
a principle is close related with possibility of cloning in quantum case.}
process with initial particle and products of decay
are staying in rest we get paradoxical situation, that linear cloning
\Eq{linclon} contradicts to principle of covariance, or more simply,
uniformity of space, because we have chosen axis in space, {\em i.e.} the
two opposite directions of spin corresponding basis\footnote{These arguments
do not have much with real physics and used to emphasize problems of
{\em given abstract mathematical model}.} $\ket{0}$ and $\ket{1}$ that
only could be cloned and so this ``space axis'' could be found after
numerous repeating of same experiment with different states.

Contrary, the nonlinear evolution like \Eq{clon12} does not have such a
problem. It is clear also that such arguments may not be applied to
apparently equal scheme \Eq{clon22}, because there is ancillary
system in state $\ket{0}$ corresponding to certain direction of spin
and so here is not necessary to introduce some non-homogeneity in model
of physical interaction, because a prefered direction presents in initial
conditions\footnote{In example with laser it corresponds to a question:
``How does it possible to clone only two fixed state of polarization ---
it breaks rotational symmetry?'' and suggestion: ``This symmetry may be
broken due to spins and nonzero angular momenta of atoms and electrons
involved in stimulated emission''.}.

The principle of covariance is very significant and universal, because
it is related with inevitable conditions, like independence of physical
laws from coordinate system used by us for its description and example above
demonstrates such kind of relations. On the other hand, the linearity
and covariance is particular case of the same mathematical idea: let us
consider some object with operation `$\star$', then {\em homomorphism} is map
$H$ to other object with property:
\begin{equation}
 H(a \star b) = H(a) \star H(b).
\end{equation}
It is clear, that if `$\star$' is addition, then ``homomorphism'' is
``linearity'', but if `$\star$' is operation of composition in group of
transformation of space-time, then ``homomorphism'' is ``covariance''.

Using formal mathematical language, principle of covariance can be expressed
as homomorphism between transformations of wave vector and symmetries of
space \cite{QFT}. So, if quantum evolution could accept some fundamental
processes between Hilbert spaces with different dimensions like \Eq{evol12},
then general physical principles rather would forget linear evolution, than
cloning.

But here again the problem \cite{cloname} with instantaneous (``superluminal'')
communications should be considered with necessary care, but it is not the
subject of present note.

\section*{Acknowledgements}
I am grateful to Asher Peres for comments.

\end{document}